\documentclass[11pt,aps,showpacs,preprintnumbers,amsmath,amssymb,nofootinbib,preprint]{article}

\usepackage{epsfig}
\usepackage{dcolumn}
\usepackage{amsmath}
\usepackage{color}
\usepackage{graphicx}

\newcommand{\ba} {\begin{eqnarray}}
\newcommand{\ea} {\end{eqnarray}}
\def \be  {\begin{equation}}
\def \ee  {\end{equation}}
\def \ee  {\end{equation}}
\def \bea {\begin{eqnarray}}
\def \eea {\end{eqnarray}}

\newcommand{\nn}{\nonumber}

\renewcommand{\figurename}{{\bf Fig.}}
\renewcommand{\tablename}{{\bf Tab.}}
\usepackage{graphicx}
\usepackage{dcolumn} 
\usepackage{bm}      
\begin{document}

\renewcommand{\figurename}{Fig.}
\renewcommand{\tablename}{Tab.}
\title{\vspace*{-1cm}\hfill {\small\bf ECTP-2014-09} \\ \hfill {\small\bf WLCAPP-2014-09}\\
\vspace*{1cm}
FLRW Cosmology with Horava-Lifshitz Gravity: Impacts of Equations of State}

\author{A.~Tawfik$^{1,2}$\footnote{Corresponding author: a.tawfik@eng.mti.edu.eg} 
~and~E. Abou El Dahab$^{1,3}$  \\
 {\small $^1$Egyptian Center for Theoretical Physics (ECTP), Modern  University for Technology and Information (MTI),}\\
 {\small 11572 Cairo, Egypt}\\
 {\small $^2$World Laboratory for Cosmology And Particle Physics (WLCAPP), Cairo, Egypt}  \\
 {\small $^3$Faculty of Computers and Information, Modern University for Technology and Information (MTI), 11671 Cairo, Egypt}
}

\date{}
\maketitle

\begin{abstract}

Inspired by Lifshitz theory for quantum critical phenomena in condensed matter, Horava proposed a theory for quantum gravity with an anisotropic scaling in ultraviolet. In Horava-Lifshitz gravity (HLG), we have studied the impacts of six types of equations of state on the evolution of various cosmological parameters such as Hubble parameters and scale factor. From the comparison of the general relativity gravity with the HLG with detailed and without with non-detailed balance conditions, remarkable differences are found. Also, a noticeable dependence of singular and non-singular Big Bang on the equations of state is observed. We conclude that HLG explains various epochs in the early universe and might be able to reproduce the entire cosmic history with and without singular Big Bang.
 
\end{abstract}


\maketitle

\section{Introduction}

The perturbative nonrenormalizability of the theory of general relativity (GR) is one of the greatest obstacles against proposing quantum gravity theory that should be correct at all scales. The latter belongs to the yet-unsolved problems in the fundamental physics \cite{carlip}. Weakness of the gravitational force is another difficulty preventing the quantum effects of gravity to be detectable at the currently-probed scales. The recent detection of the gravitational waves is strictly limited to pulsars in binary star systems \cite{LIGO2015}. Accordingly, the resulting gravitational waves become very great.

At a Gaussian fixed point, GR does not satisfy the perturbative renormalizability because the gravitational coupling ($G$) is a dimensioned quantity ($G\simeq \text{mass}^{-2}$) and accordingly the ultraviolet (UV) divergences are uncontrollable. By introducing higher terms of the scalar curvature, the action shall be cured. On the other hand, this introduces an additional problem, for instance, to the uniqueness \cite{thesisF}. From a minimal approach that will be elaborated below, which assures perturbatively renormalizable UV completion of GR as quantum field theory (QFT) for $4$-dimensional metric field, one looses some properties of GR and creates additional degrees-of-freedom. Such an approach benefits from the success of QFT in describing all forces (except gravity so-far) and the standard model of the elementary particles. The scalar field theory, that was proposed by Lifshitz in order to explain the quantum critical phenomena in condensed matter \cite{lifshitz}, inspired Horava to propose a theory for the quantum gravity with an anisotropic scaling in UV. This new approach is thus known as Horava-Lifshitz gravity (HLG) \cite{horava2009a,HLGRef1,HLGRef2,HLGRef3,HLGRef4,HLGRef5,HLGRef6,HLGRef7,HLGRef8,HLGRef9}, where the dynamical critical exponent ($z$) makes the theory non-relativistic.

In analogy to critical systems, the UV fixed point has been treated \cite{hlg1}. The HLG  approach assures causal dynamical triangulations \cite{amb1}, renormalization group approaches based on asymptotic safety \cite{FP1,FP2} and symmetries of GR. This approach was discussed in an extensive amount of literature \cite{refrns1,refrns2,refrns3}, where problems such as, internal consistency and compatibility with observations are treated, as well. The new dimensionless coupling $\lambda$ is assumed to approach unity in the infrared (IR) limit with the argumentation that at the HLG-charactering parameter $\lambda\neq1$, HLG theory can't be reduced to GR. For the sake of completeness, we mention that this is not always the case for all HLG models \cite{disser}.

Recent astrophysical observations obviously indicate that the universe expands with an accelerating rate, probably due to dark energy \cite{expUni,Tawfik:2014dza}. Besides inflation, a series of symmetry-breaking phase transitions, where topological defects may have been formed \cite{ir1}, were conjecture to take place. On the other hand, the HLG-type approach \cite{horava2009a} proposes that the modification in gravity is responsible for the accelerated expansion of the universe. 

Furthermore, various modified theories for gravity can be implemented \cite{nori,masud}. They successfully unify various inflation models with the late-time acceleration and cosmological observations. But when endorsing any proposal about (non)singular Big Bang due to controversial quantum gravity approaches such as generalized uncertainty principle (GUP) and modified dispersion relation (MDR) \cite{Tawfik:2014zca} should be seen as unwarranted claims. the Big Bang singularity is a fundamental problem, which should be tackled by a good quantum gravity theory, such as HLG or even better.

The present paper is organized as follows. The cosmological evolution governed by a simple version of HLG shall be introduced. The Horava-Lifshitz gravity is introduced in section \ref{sec:HLG}. The field equations in Friedmann-Lemaitre-Robertson-Walker (FLRW) cosmology are elaborated in section \ref{sec:fe}. Section \ref{sec:nondet} and section \ref{sec:det} are devoted to  Horava-Lifshitz gravity with non-detailed and detailed balance conditions, respectively. For the different equations-of-state (EoS), different solutions for the resulting differential equations shall be outlined. An extensive comparison between scale factor in cosmological radiation, matter, $\Lambda$CDM, de Sitter, Chaplygin gas \cite{chplyginREF} and quantum chromodynamical (QCD) EoS shall be introduced. The results and discussion are given in section \ref{sec:disc}. To section \ref{sec:conc} the final conclusions are assigned.

\section{Reminder of Horava-Lifshitz gravity}
\label{sec:HLG}

The basic assumptions of Horava-Lifshitz gravity are antisotropic UV scalings between Minkowskian space and time 
\begin{eqnarray}
t \longrightarrow  l^{z}\, t, \qquad & &
x^{i} \longrightarrow  l\, x^{i},
\end{eqnarray}
where $z$ is the dynamical critical exponent given in UV, $l$ is constant by which the scaling is performed and to the critical exponent the value $2$ is assigned. Because the anisotropic scaling implies a preferred time coordinate, the usual $4$-dimensional Lorentzian metric can't be the only fundamental structure on a pseudo-Riemannian (Lorentzian) manifold. 

The original theory postulated $z$ to take the value assuring dimensionless gravitational coupling, which in turn guarantees power-counting renormalizability. At $z=1$, the symmetries of GR are recovered. In order to single out a spacial (temporal) coordinate in a differentiable manifold, a codimension foliation on the manifold as a basic structure of the theory was proposed  \cite{horava2009a}. The foliation structure fulfils local Galilean invariance but implies impossible full-diffeomorphism invariance in GR. The latter means the existence of an undo one-to-one mapping $f$ and its inverse besides their differentiabilities. Thus, GR may be considered as an emerging in an infrared fixed point \cite{ir1}. 

The HLG is a projectable approach minimizing the number of independent couplings in the potential and adopting an extra principle to contract the potential, i.e. detailed balance conditions. In constructing the theory, a new set of symmetries should be introduced. Imposing invariance under foliation-preserving diffeomorphisms, $\delta\, x^i=\xi (x^i, t)$ and $\delta\, t=f(t)$ would define field content, which paradoxically not always the one in GR and would make the theory violates the Lorentz invariance principle. On the other hand, for Arnowitt, Deser and Misner [ADM] formalism \cite{ADMref} (GR with global time foliation, $3$-dimensional metric of the spatial hypersurfaces gives all the dynamics), the metric is given as
\begin{eqnarray}
ds^{2} &=& -\, N^{2}\, dt^{2} + g_{ij}\, (dx^{i} + N^{i}\, dt)(dx^{j} + N^{j}\, dt), \label{ADM1}
\end{eqnarray}
where $i,\, j=1,2,3$, and $N\,$ ($N_{i}$) being lapse function (shift $3$-vector), which are the gauge fields of the diffeomorphism group. The lapse variable is taken to be just time-dependent so that the projectability condition holds and by using the foliation-preserving difeomorphisms $N$ can be fixed to unity. 

The variables in Eq. (\ref{ADM1}) are dynamical so that under the above scaling 
\begin{eqnarray}
N \longrightarrow  N, \qquad
g_{ij} \longrightarrow g_{ij},  && 
N_{i} \longrightarrow  l^{2}\, N_{i}, \qquad
N^{i} \longrightarrow  l^{-2}\, N^{i}.
\end{eqnarray}
In terms of the metric Eq. (\ref{ADM1}), the Ricci scalar reads
\begin{eqnarray}
R &=&K_{i j}\, K^{i j} - K^{2} + R^{(3)} + 2 \nabla_{\mu}\left(n^{\mu}\, \nabla_{\nu}\, n^{\nu} - n^{\nu}\, \nabla_{\nu}\, n^{\mu}\right),
\end{eqnarray}
where $K=g^{i\,j}\, K_{i\,j}$, $K_{i\,j}=(\partial_t\, g_{i\,j}^{(3)} - \nabla_i^{(3)}\, N_j - \nabla_j^{(3)}\, N_i)/(2 N)$ is the extrinsic curvature of spatial slices, $ R^{(3)}$ is the spacial scalar curvature, and $n^{\nu}$ is a unit vector perpendicular to a hypersurface of constant time. $\nabla$ is the covariant derivative on the spatial slice.

The HLG action is decomposed of kinetic, potential and matter parts.
%
\begin{itemize}
\item The kinetic action can be built with at most two time-derivatives of the metric. This is a universal and most general invariant term. The new dimensionless coupling $\lambda$, whose role in the theory is still a subject of debate, differentiates the HLG kinetic term from the GR one 
\begin{eqnarray}
S_{k} = \int dt\, dx^{3}\, 2\, \dfrac{\sqrt{g}\, N}{k^{2}} \left(K_{ij}\, K^{ij}- \lambda K^{2}\right).
\end{eqnarray}
When combining this kinetic term with the line element and the anisotropic scaling of space and time, the scaling dimension of the gravitational coupling can be determined. 

\item The potential action term \cite{SIX1,SIX2} (GR contains only two derivatives) is affected by the choice between projectable and non-projectable theories \cite{disser} 
\begin{eqnarray}
S_{v}&=& - \frac{2}{\kappa^2}\int dt d^{3} x \sqrt{g} N \left[\sigma + \gamma\, R + \gamma_{1}\, R^{2} + \gamma_{2}\, R_{ij}\, R^{ij} + \xi\,  \varepsilon^{ijk}\, R_{il}\, \nabla_{j}\, R^{l}_{k}   \right. \\ \nonumber
&+& \left. \sigma_{1}\, R^{3} + \sigma_{2}\, R\, R^{ij}\, R_{ij} + \sigma_{3}\, R^{j}_{i}\, R^{k}_{j}\, R^{i}_{k}
 + \sigma_{4}\, \nabla_{i}\, R\, \nabla^{i}\, R + \sigma_{5}\, \nabla_{i}\, R_{jk}\, \nabla^{i}\, R^{jk} \right]. \label{eq:termP}
\end{eqnarray}
The coefficients are given in just three parameters, $\mu$, $\Lambda_{\omega}$ and $\zeta$ \cite{HL1} 
\begin{eqnarray}
\sigma = -\frac{3\kappa^{4}\mu^{2} \Lambda_{\omega}^{2}}{16(3\lambda-1)}, \qquad 
\gamma=- 3\, \Lambda_{\omega}\, \sigma,  & & 
\gamma_{1}=\frac{\kappa^{4}\mu^{2}(1-4\lambda)}{64(3\lambda-1)}, \qquad
\gamma_{2}=\frac{\kappa^{4}\mu^{2}}{16}, \qquad
\xi=-\frac{64}{\mu\, \zeta^{2}}\, \gamma_2, \nonumber \\
\sigma_{1}=\frac{\kappa^{4}}{8\zeta^{4}}, \qquad 
\sigma_{2}=-5\, \sigma_1, &&
\sigma_{3}=6\, \sigma_1, \qquad
\sigma_{4}=-\frac{3}{4}\,\sigma_1, \qquad 
\sigma_{5}=2\, \sigma_1, \label{eq:variables}
\end{eqnarray}
where $\kappa^{2}=8\pi G/c^{4}$ and $\Lambda_{\omega}$ are Einstein coupling and cosmological constant, respectively. The lower-order variables in expression (\ref{eq:termP}); the cosmological constant and the Ricci scalar, match the ones in GR. 

\item The matter part of the action \cite{matter1} is given as
\begin{eqnarray}
 S_{m}=\int dt\, dx^{3}\, \sqrt{g}\, N\, L_{m},
\end{eqnarray}
where $L_{m}$ is the Lagrangian density of matter fields $L_{m}(N,N_{i},g_{ij},\phi)$. 
\end{itemize}

\subsection{Scale-invariant quantum fluctuations}

In Ref. \cite{0904.2190}, a simple scenario for the scale-invariant quantum fluctuations was introduced. This was based on HLG without an additional scalar degree-of-freedom. Accordingly, it is believed that the inflation is self-consistently existing in the early universe. For this simple scenario, the detailed balance conditions were not necessary but the inflation scenario itself may be still or no longer needed, for instance, serious horizon problems remain unsolved, such as monopole and domain walls. They still require inflation with slow-roll conditions. 

Later on many authors seem to disagree with Ref. \cite{0904.2190}. The inflation was studied in HLG without the projectability conditions \cite{1208.2491}. But, opposite to the proposal of Ref \cite{0904.2190}, the linear scalar perturbations equations of the FLRW universe are derived for a single scalar field. A master equation of the perturbations has been specified for a particular gauge. The power spectrum and the spectrum index of the comoving curvature perturbations have been determined. It was noticed that the perturbations remain scale-invariance, and HLG without the projectability conditions is consistent with all current cosmological observations. This is another solid support for adding scalar field(s).

In framework of nonrelativistic HLG with the projectability conditions and an arbitrary coupling constant $\lambda$, the inflation has been studied \cite{1201.4630}. Accordingly, the FLRW universe without specifying the gauge is necessarily flat. But by adding a single scalar field, it was noticed that both metric and scalar field become strongly coupled and almost identically oscillating in sub-horizon regions. In super-horizon regions, the comoving curvature perturbation remains constant although the FLRW perturbations become adiabatic. Furthermore, the perturbations of the slow-roll parameters, for instance, both scalar and tensor are found scale-invariant. Concrete tuning the coupling coefficients makes the spectrum index of the tensor perturbation identical as that in GR. But the ratio of scalar to tensor spectra can be similar to that from GR and seems to depend on the spatial higher-order derivative terms.

Furthermore, the emergence of finite-time future singularities has been studied in Ref. \cite{1006.3387}. It was found that such singularities can be cured by adding a higher-order spatial derivative term. Recently, the general formulas for the inflationary power spectra of scalar and tensor are driven in the presence of a scalar field \cite{1409.1984}.

We conclude that in simple scenarios for scale-invariant cosmological perturbations with HLG, the inflation is not necessarily guaranteed.

\section{Field equations in FLRW cosmology}
\label{sec:fe}

For a cosmological context, we recall FLRW metric, which is an exact solution of the Einstein's field equations of GR, 
\begin{eqnarray}
N=1, \qquad  N^{i}=0, \qquad g_{ij}=a^{2}(t) \gamma_{ij},
\end{eqnarray}
where $a(t)$ is the scale factor and $\gamma_{ij}dx^{i}dx^{j}=dr^{2}/(1-kr^{2})+r^{2} d\Omega_{2}^{2}$.
Accordingly, the homogeneous and isotropic metric is given as
\begin{eqnarray}
ds^{2}=-dt^{2}+a(t)^{2}\left[\frac{dr^{2}}{1-k r^{2}}+r^{2}(d\theta^{2}+\sin^{2}\theta\, d\phi^{2})\right],
\end{eqnarray}
with the curvature constant $k=-1$, $0$ and $1$ represents open, flat or closed universe, respectively. 

\subsection{Horava-Lifshitz gravity with non-detailed balance conditions}
\label{sec:nondet}

In  HLG with non-detailed balance conditions and by varying $N$ and $g_{ij}$, the Friedman equations can be extracted \cite{ir1},
\begin{eqnarray}
3(3\lambda - 1) H^{2} &=& \frac{\kappa^{2}}{2} \rho + 6 \left[\frac{\sigma}{6} + K \frac{\gamma}{a^{2}} + 
2 K^{2} \frac{3 \gamma_{1} + \gamma_{2}}{a^{4}} + 4 K^{3} \frac{9 \sigma_{1} + 3\sigma_{2} + \sigma_{3}}{a^{6}}\right], \label{13} \\
(3 \lambda - 1) \left(\dot{H} + \frac{3}{2} H^{2}\right) &=& - \frac{\kappa^{4}}{4} p - 3 \left[-\frac{\sigma}{6} - K \frac{\gamma}{3 a^{2}}+ 2 K^{3} \frac{9\, \sigma_{1} + 3\, \sigma_{2} + \sigma_{3}}{a^{6}}\right]. \label{14}
\end{eqnarray}
By substituting with the variables given in  Eqs. (\ref{eq:variables}), 
%
 (\ref{13}) and (\ref{14}), the Freidmann equations read
\begin{eqnarray}
H^2 &=& \dfrac{\kappa^2}{6 (3 \lambda - 1)} \rho - \dfrac{\kappa^4 \mu^2 \Lambda^2_w}{16 (3 \lambda  - 1)^2} + \dfrac{\kappa^4 K \mu^2 \Lambda_{\omega}}{8 (3 \lambda - 1)^2 a^2 } - \dfrac{\kappa^4 K^2  \mu^2}{16 (3 \lambda - 1)^2 a^4}, \label{19}\\
\dot{H}+ \dfrac{3}{2} H^2 &=& - \dfrac{\kappa^2}{4 (3 \lambda - 1)} p - \dfrac{3\, \kappa^4\, \mu^2\, \Lambda^2_w}{32 (3\, \lambda - 1)^2} + \dfrac{K\, \kappa^4\, \mu^2 \, \Lambda_{\omega}}{16 (3\, \lambda -1)^2\; a^2}.\label{20}
\eea
where $H=\dot{a}/a$, $\dot{H}=\ddot{a}/a - H^2$ and $\ddot{a}/a = \dot{H} + H^{2}$. Then, from Eqs. (\ref{19}) and (\ref{20}), we get
\begin{eqnarray}
\ddot a = \left\{- \dfrac{\kappa^2}{12 (3 \lambda -1)} [\rho+3 p]  - \dfrac{\kappa^4\, \mu^2\, \Lambda_{\omega}^2}{16(3 \lambda -1)^2}\right\}\; a + \dfrac{K^2\, \kappa^4\, \mu^2}{32(3 \lambda - 1)^2} a^{-3}. \label{21b}
\end{eqnarray}

Based on equations of state (EoS) describing the cosmological background geometry, this differential equation can be solved. From the continuity equation 
\begin{eqnarray}
\dot{\rho} + 3 H (\rho + p) &=& 0, \label{eq:rhodot}
\end{eqnarray}
and with the EoS $p=\omega\, \rho$,  we get $\rho=\rho_0\, a^{-3(1+\omega)}$ and $p=\omega\, \rho_0\, a^{-3(1+\omega)}$, where $\omega$ - in natural units - is a dimensionless quantity. It is related to the speed of sound squared within the system of interest. Then, Eq. (\ref{21b}) can be expressed as
\begin{eqnarray}
\ddot a = - \dfrac{\kappa^2\, \rho_0(1+3\omega)}{12 (3 \lambda -1)} \; a^{-2-3\omega} - \dfrac{\kappa^4 \mu^2 \Lambda_{\omega}^2}{16(3 \lambda -1)^2}\; a + \dfrac{K^2 \kappa^4 \mu^2}{32(3 \lambda - 1)^2} a^{-3}. \label{21a}
\end{eqnarray}

\subsubsection{Perturbation Stability}

Equation (\ref{20}) can be rewritten as 
\bea
\dot{H} &=& - \frac{2}{3} H^2 - \frac{\kappa^2}{4 (3 \lambda -1)} p - L + \frac{M}{a^2}, \label{20Pert}
\eea
where $L=3 \kappa^4 \mu^2 \Lambda_{\omega}^2/[32 (3 \lambda -1)^2]$ and $M=K \kappa^4 \mu^2 \Lambda_{\omega}/[16 (3 \lambda -1)]$. Assuming perturbation in Eq. (\ref{20Pert}) leads to
\bea
\dot{H} + \dot{\delta} H &=& - \frac{2}{3} H^2 - 3 H \delta\, H- \frac{\kappa^2}{4 (3 \lambda -1)} p - L + \frac{M}{a^{2}} - 2 \frac{M}{a^2}\, \frac{\delta a}{a}. \label{20Pert2}
\eea
For an EoS in which $\delta p \ll \delta \rho$ and for the first-order perturbation, we get
\bea
\dot{\delta} H &=& - 3 H \delta H - 2 \frac{M}{a^2}\, \frac{\delta a}{a}. \label{20Pert3}
\eea
However from the continuity equation, Eq. (\ref{eq:rhodot}), 
\bea
\delta & \equiv & \frac{\delta \rho}{\rho} = -3 (1+\omega) \frac{\delta a}{a}.
\eea
Then 
\bea
\dot{\delta}\, H &=& {\cal M}\, \delta\, H + {\cal N}\, \delta, \\
\dot{\delta} &=& {\cal P}\, \delta\, H + {\cal Q}\, \delta,
\eea
with ${\cal M}= -3 H$,
${\cal N}= 2 M / [3 (1+\omega) a^2]$,
${\cal P}= -3 (1+\omega)$, and
${\cal Q}= -3 H \omega$. Thus,
\bea
\ddot{\delta}\, H - \left({\cal M} +\frac{\dot{{\cal N}}}{{\cal N}} + {\cal Q}\right) \dot{\delta}\, H 
- \left(\dot{{\cal M}} - {\cal M} \frac{\dot{{\cal N}}}{{\cal N}} + {\cal P} {\cal N} - {\cal Q} {\cal M}\right) \delta\, H &=& 0,
\eea
For $\dot{{\cal N}}/{\cal N}=-2 \dot{a}{a} = -2 H$, the term with the first-order perturbation reads $(5+3 \omega) H$ is apparently positive. This gives a stable solution irrespective to the third term in left-hand side.

\subsubsection{Various equations-of-state}

In describing the cosmic geometry, we wanted to analyse systematically the effects of implementing various equations-of-state.
\begin{itemize} 

\item  When the cosmological geometry is dominantly filled with matter (dust approximation), $\omega=0$, Eq. (\ref{19}) can be rewritten as
 
\begin{eqnarray}
\dot{a}^2 =\dfrac{\kappa^2 \rho_{\circ}}{6(3 \lambda -1)} a^{-1} - \dfrac{\kappa^4 \mu^2 \Lambda_w^2}{16(3 \lambda -1)^2} a^2 - \dfrac{K^2 \kappa^4 \mu^2}{16(3 \lambda - 1)^2} a^{-2} + \dfrac{\kappa^4 K \mu^2 \Lambda_w^2}{8(3 \lambda - 1)^2}. \label{eq:add0}
\end{eqnarray}
When assuming that the cosmological constant $\Lambda_{\omega}$ and Einstein coupling $\kappa$ are negligibly small, Eq. (\ref{eq:add0}) gets the solution
\begin{eqnarray}
t &=& \frac{2}{3} (A_1\, a - 2\, B_1) A_1^{-2}\, \sqrt{A_1\, a+B_1}. \label{eq:add0a}
\end{eqnarray}
where $A_1=\kappa^2 \rho_{\circ}/[6(3 \lambda -1)]$ and $B_1= - K^2 \kappa^4 \mu^2/[16(3 \lambda - 1)^2]$. Deducing scale factor does not restrict the applicability of the present analysis. Other cosmological quantities such as Hubble and deceleration parameters are good examples.

\item For cosmological geometry filled with radiation (perfect fluid approximation), $\omega=1/3$, Eq. (\ref{19}) becomes
\begin{eqnarray}
\dot{a}^2 &=& \left(\dfrac{\kappa^2\, \rho_0}{6 (3 \lambda -1)} - \dfrac{K^2 \kappa^4 \mu^2}{16(3 \lambda - 1)^2}\right)\; a^{-2} - \dfrac{\kappa^4 \mu^2 \Lambda_w^2}{16(3 \lambda -1)^2}\; a^2 + \dfrac{\kappa^4 K \mu^2 \Lambda_w^2}{8(3 \lambda - 1)^2}, \label{solsim}
\end{eqnarray}
which has the following solution
\begin{eqnarray}
t = \frac{1}{\sqrt{-B_2}} \arcsin \left( \dfrac{-2 B_2 a^2 - C_1}{\sqrt{-4 B_2 A_2 + C_1^2}} \right), \label{eq:ndwp13}
\end{eqnarray}
where  $A_2=\kappa^2\, \rho_0/[6 (3 \lambda -1)] - K^2 \kappa^4 \mu^2/[32(3 \lambda - 1)^2]$ and $B_2=-\kappa^4 \mu^2 \Lambda_w^2/[16(3 \lambda -1)^2]$ and $C_1=\dfrac{\kappa^4 K \mu^2 \Lambda_w^2}{8(3 \lambda - 1)^2}$.

\item For cosmological geometry filled with cold dark-matter, $\omega=-1/3$, Eq. (\ref{19}) is written as,
\begin{eqnarray}
\dot{a}^2 = \left(\dfrac{\kappa^2 \rho_{\circ}}{6(3 \lambda -1)} + \dfrac{\kappa^4 K \mu^2 \Lambda_w^2}{8(3 \lambda -1)^2}\right)\, -  \dfrac{K^2 \kappa^4 \mu^2}{16(3 \lambda - 1)^2} a^{-2} - \dfrac{\kappa^4 \mu^2 \Lambda_w^2}{16(3 \lambda - 1)^2} a^2,
\end{eqnarray}
The solution looks similar to the one in Eq. (\ref{solsim})
\begin{eqnarray}
t = \frac{1}{\sqrt{-B_2}} \arcsin \left( \dfrac{-2 B_2 a^2 +B_2 - C_1}{\sqrt{-4 B_1 B_2 + (-B_2+C_1)^2}} \right).
\end{eqnarray} 

\item For cosmological geometry filled with vacuum (dark) energy (cosmological constant), $\omega=-1$, then Eq. (\ref{19}) can be written as \cite{soluts},
\begin{eqnarray}
\dot{a}^2 = \left(\dfrac{\kappa^2 \rho_{\circ}}{6(3 \lambda -1)} - \dfrac{\kappa^4 \mu^2 \Lambda_w^2}{16(3 \lambda -1)^2}\right)\, a^2 -  \dfrac{K^2 \kappa^4 \mu^2}{16(3 \lambda - 1)^2} a^{-2} + \dfrac{\kappa^4 K \mu^2 \Lambda_w^2}{8(3 \lambda - 1)^2} \label{eq:awm1}
\end{eqnarray}
Then, the solution reads
\begin{eqnarray}
t = \frac{1}{\sqrt{A_2+B_1}} \arcsin \left(\dfrac{2(A_1+B_2)a^2-C_1}{\sqrt{4(A_1+B_2)(B_1)+C_1^2}} \right),
\end{eqnarray}

\item For cosmological geometry filled with Chaplygin gas \cite{chplyginREF}, which is a hypothetical (non-baryonic) substance satisfying $p=-A/\rho^{\alpha}$, where  $A$ is positive and $\alpha$ ranges between $0$ and $1$, i.e. negative pressure, and does not cluster, gravitationally. It was named after Sergey Chaplygin (1869-1942), and expresses energy of quantum vacuum, and gravity modification. It might signal extra-dimensions and is conjectured as a candidate for dark energy. The energy density and pressure, respectively, read
\begin{eqnarray}
\rho &=& \left[A+\frac{B}{a^{3(1+\alpha)}}\right]^{1/(1+\alpha)}, \label{eq:chplp}\\
p &=&-A\, \left[A+\frac{B}{a^{3(1+\alpha)}}\right]^{-\alpha/(1+\alpha)}, \label{eq:chple}
\end{eqnarray}
where $B$ is an integration constant. 

In FLRW cosmology, the Chaplygin equations of state, Eqs. (\ref{eq:chplp}) and (\ref{eq:chple}), are substituted in Eqs. (\ref{19}) and (\ref{21b}) to give
\begin{eqnarray}
\ddot{a} &=& - \frac{\kappa^2\, A^{\frac{1}{1+\alpha}}}{12(3\lambda-1)} \left\{\left[\left(1+\frac{B/A}{1+\alpha} a^{-3(1+\alpha)}+\cdots\right) - 3  \left(1+\frac{\alpha B/A}{1+\alpha}\, a^{-3(1+\alpha)}+\cdots\right)\right] \right.\nn\\
&& \left. - \dfrac{\kappa^4\, \mu^2\, \Lambda_\omega^2}{16(3\lambda-1)^2}\right\}\, a + \dfrac{K^2\, \kappa^4\, \mu^2}{32(3\lambda-1)^2}\, a^{-3}. \label{eq:daachapA0}
\end{eqnarray}

When ignoring the Taylor terms with orders $\geq 2$, then
\begin{eqnarray}
\dot{a}^2 &=& \left( \frac{\kappa^2\, A^{\frac{1}{1+\alpha}}}{6(3\lambda-1)} - \frac{\kappa^4\, \mu^2\, \Lambda_\omega^2 }{16(3\lambda-1)^2} \right) a^2 + \frac{\kappa^2 B A^{\frac{-\alpha}{1+\alpha}}}{6(3 \lambda - 1)(1+\alpha)}\, a^{-(1+3 \alpha)}-\dfrac{K^2\, \kappa^4\, \mu^2 }{16(3\lambda-1)^2}\, a^{-2} \nn\\
&+& \dfrac{\kappa^4\, K \mu^2 \Lambda_\omega^2 }{8(3 \lambda - 1)^2}. \label{eq:daachapA1} 
\end{eqnarray}

The possible solution for this differential equations depends on the choices of $A$ and $\alpha$. Here, we elaborate different cases:
\begin{itemize}
\item First for a modified Chaplygin gas \cite{chplyginREF} with $A\neq 1$ and by assuming that $\alpha=1/3$, then Eq. (\ref{eq:daachapA1}) is reduced to
\begin{eqnarray}
\ddot{a} &=& \left\{\dfrac{- \kappa^2\, A^{3/4}}{12(3\lambda-1)} \left[-2 +\frac{3}{2}\, \frac{B}{A}\, a^{-4}\right] - \dfrac{\kappa^4\, \mu^2\, \Lambda_{\omega}^2}{16(3\, \lambda-1)^2}\right\}\, a + \dfrac{K^2\, \kappa^4\, \mu^2}{32(3\lambda-1)^2}\, a^{-3}, \label{eq:addChapB1} \\
\dot{a}^2 &=& \left[\dfrac{\kappa^2\, A^{3/4}}{6(3\lambda-1)} - \dfrac{\kappa^4\, \mu^2\, \Lambda_{\omega}^2}{16(3\, \lambda-1)^2}\right] a^2 + \left[\dfrac{\kappa^2\, B A^{-1/4}}{8(3\lambda-1)} - \dfrac{K^2\, \kappa^4\, \mu^2}{16(3\lambda-1)^2} \right] a^{-2} \nn\\
&+& \dfrac{\kappa^4\, K \mu^2 \Lambda_{\omega}^2}{8 (3 \lambda - 1)^2},
\end{eqnarray}
which has a solution
\begin{eqnarray}
t &=&\frac{1}{2\sqrt{A_3}} \left[\ln \left(a^2+ \dfrac{C_1}{2A_3}+ \sqrt{a^4+\dfrac{B_3}{A_3} +\dfrac{C_1}{A_3}a^2} \right) \right],
\end{eqnarray}
where
\begin{eqnarray}
A_3 &=& \left[\dfrac{\kappa^2\, A^{3/4}}{6(3\lambda-1)} - \dfrac{\kappa^4\, \mu^2\, \Lambda_{\omega}^2}{16(3\, \lambda-1)^2}\right], \nn \\
B_3 &=& \left[\dfrac{\kappa^2\, B A^{-1/4}}{8(3\lambda-1)} - \dfrac{K^2\, \kappa^4\, \mu^2}{16(3\lambda-1)^2} \right]. \nn
\end{eqnarray}

\item For matter, the pressure vanishes or $A=0$. Eq. (\ref{eq:daachapA1}) is reduced to
\begin{eqnarray}
\dot{a}^2 = - \dfrac{\kappa^4\, \mu^2\, \Lambda_{\omega}^2}{16(3\, \lambda-1)^2}\, a^2 - \dfrac{K^2\, \kappa^4\, \mu^2}{16(3\lambda-1)^2}\, a^{-2} + \dfrac{\kappa^4\, K \mu^2\ \Lambda_{\omega}^2}{8(3 \lambda - 1)^2},
\end{eqnarray}
which has a solution
\begin{eqnarray}
t = \frac{1}{\sqrt{-B_1}} \arcsin \left(\dfrac{ - 2 B_1 a^2 - C_1}{\sqrt{-4 B_1 B_2 + C_1^2}} \right).
\end{eqnarray}

\end{itemize}

\item For cosmological geometry filled with QCD matter, the EoS can be deduced from the recent lattice QCD simulations \cite{fodor2013}, where $w$ ranges between $\sim 1/6$ and $\sim 1/4$. For sake for completeness, we highlight the importance of precise estimation of EoS from heavy-ion collisions \cite{Tawfik:2015kwa}. By substituting the deduced values of $\omega$ in Eq. (\ref{21a}), the resulting differential equation turns to be unsolvable. For simplicity, the first term in Eq. (\ref{21a})  can be approximated,  
\begin{eqnarray}
\ddot{a} &\approx & \left[- \frac{3}{4}\, \frac{\kappa^2\, \rho_0}{12 (3 \lambda -1)} + \dfrac{K^2 \kappa^4 \mu^2}{32(3 \lambda - 1)^2}\right]\; a^{-3} - \dfrac{\kappa^4 \mu^2 \Lambda_w^2}{16(3 \lambda -1)^2}\; a, \\
\dot{a}^2 &\approx & \left[ \frac{\kappa^2\, \rho_0}{16 (3 \lambda -1)} - \dfrac{K^2 \kappa^4 \mu^2}{32(3 \lambda - 1)^2}\right]\; a^{-2} - \dfrac{\kappa^4 \mu^2 \Lambda_w^2}{16(3 \lambda -1)^2}\; a^2. 
\end{eqnarray}
Similar to Eq. (\ref{eq:awm1}), the solution reads
\begin{eqnarray}
t &=&\frac{{\cal A}\, \log\left[2\left(\sqrt{B_2}\; {\cal A} + B_2\, a^2\right)\right]}{2\, \sqrt{B_2}\; {\cal A}},
\end{eqnarray}
where ${\cal A}=\left[B_2\, a^4 + (A_1+B_1)\right]^{1/2}$. 

\end{itemize}

\subsection{Horava-Lifshitz gravity with detailed balance conditions}
\label{sec:det}

The detailed balance condition is a technical trick proposed to reduce the couplings. Furthermore, it was argued that HLG with detailed balance conditions has to be broken in order to enable the theory to be compatible with the observations \cite{0905.2579}. But, according to Horava's remarks, its restrictive power remains useful \cite{Horava2011a}. There is a possible connection between detailed balance conditions and the entropic origin of gravity \cite{Verlinde2011a}.  The detailed balance condition is often believed to be abandoned with the aim to obtain an ultraviolet stable scalar field in the theory. But because of its several attractive features, we wanted it to be implemented in constructing the potential instead of effective field theory. Many authors wanted to improve it by adding extra term to the action of the theory that softly violates the detailed balance conditions \cite{attempt1} as an attempt to assure a more realistic theory in its infrared-limit.  

It was shown \cite{0904.1595} that the properties of the extra scalar degrees-of-freedom are the reasons why detailed balance conditions don't lead to GR in the infrared-limit. 

From perturbation study of detailed balance conditions \cite{0905.2579}, two different strong coupling problems have been identified. The first one explains why the GR solutions are typically not recovered. The second is not necessarily associated with detailed balance conditions but refers to breaking of diffeomorphism invariance, which is required for anisotropic scaling in UV.  

The authors of Ref. \cite{0905.2579} claimed to uncover an additional mode in the detailed balance conditions satisfying an equation of motion that is of first-order in time derivative. But, the proposed mode leads to very fast exponential instabilities at short distances and becomes strongly coupled at an extremely low cutoff scale.

For  HL gravity with detailed balance conditions, the field equations read \cite{0904.1595}
\begin{eqnarray}
H^2 &=& \dfrac{2}{3\lambda - 1} \left(\dfrac{\Lambda_{\omega}}{2} + \dfrac{8 \pi G_N}{3} \rho - \kappa\, a^{-2} + \dfrac{\kappa^2}{2 \Lambda_{\omega}} a^{-4} \right), \label{26}\\
\dfrac{\ddot{a}}{a} &=& \dfrac{2}{3\lambda - 1} \left(\dfrac{\Lambda_{\omega}}{2} - \dfrac{4 \pi G_N }{3} (\rho +3\, p) - \dfrac{K^2}{2 \Lambda_{\omega}} a^{-4} \right), \label{27}
\end{eqnarray}
The continuity equations $\rho=\rho_0 a^{-3(1+\omega)}$ can be substituted in and the Friedmann equation can be derived from the action 
\begin{eqnarray}
S = \int d t\, d^3\, X \left(\ell_\circ + \ell_1 \right),
\end{eqnarray}
where
\begin{eqnarray}
\ell_\circ &=& \sqrt{g} N \left[\dfrac{2}{\kappa^2} \left(K_{ij} K^{ij} - \lambda \kappa^2 \right) + \dfrac{\kappa^2 \mu^2 \left( \Lambda_w R - 3 \Lambda_{\omega}^{2}\right)}{8\left( 1- 3 \lambda \right)} \right], \\
\ell_1 &=& \sqrt{g} N \left[\dfrac{\kappa^2 \mu^2 (1- 4\lambda)}{32(1-3\lambda)} R^2 - \dfrac{\kappa^2}{2 w^4} \left(C_{ij} - \dfrac{\mu w^2}{2} R_{ij} \right) \left( C^{ij} - \dfrac{\mu w^2}{2} R^{ij} \right) \right], 
\end{eqnarray}
and $\lambda$, $\kappa$, $\mu$, $\omega$ and $\Lambda_{\omega}$ are constants, and $R\; (R_{ij})$ being Ricci scalar (tensor). $C_{ij}$ is the Cotton tensor, $C_{ij} = \epsilon^{ikl} \nabla_{k} (R^j_l - R \delta_l^i/4)$. When comparing HLG $\ell_\circ$ to that of GR in ADM formalism, then the speed of light and gravitational constant, respectively, reads
\begin{eqnarray}
c &=& \dfrac{\kappa^2 \mu}{4} \sqrt{\dfrac{\Lambda_{\omega}}{1-3\lambda}}, \\
G &=& \frac{\kappa^2}{32 \pi c}.
\end{eqnarray}
When assuming loss of generality, $c=1$. In solving Eqs. (\ref{26}) and (\ref{27}) four cases are likely to occur:
\begin{itemize}
\item if $\rho_\circ = 0$, and $\kappa \neq 0$
\begin{eqnarray}
a^2(t) &=& \dfrac{\kappa}{\Lambda_w} + \alpha\, \exp\left\{{2\, \sqrt{\dfrac{\Lambda_{\omega}}{3\, \lambda - 1}}\; t}\right\},
\end{eqnarray}
where $\alpha$ is constant of integration (arbitrary constant).
\item if $\rho_\circ \neq 0$, and $\kappa = 0$
\begin{eqnarray}
a^4(t) &=& -\dfrac{8 \pi G}{3}\; \rho_\circ + \alpha\,  \exp\left\{{4\, \sqrt{\dfrac{\Lambda_{\omega}}{3\, \lambda - 1}}\; t}\right\}.
\end{eqnarray}

\item if $\rho_\circ = \kappa = 0$
\begin{eqnarray}
a(t) &=&  \alpha\,  \exp\left\{\sqrt{\dfrac{\Lambda_{\omega}}{3\, \lambda -1}}\; t\right\}.
\end{eqnarray}

\item if $\rho_\circ \neq 0$, and $\kappa \neq 0$
\begin{eqnarray}
a^2(t) &=& \dfrac{\kappa}{\Lambda_{\omega}} + \alpha \exp\left\{2\, S\, \sqrt{\dfrac{\Lambda_{\omega}}{3\lambda -1}}\; t\right\}, 
\end{eqnarray}
with $S = 1 + 8\, \pi\, G\, \rho_\circ\, \Lambda_{\omega}^{2}/(3\, \kappa^2)$.

\end{itemize}

The solutions of the differential equations (\ref{26}) and (\ref{27}) depend on the EoS charactering the cosmological background geometry. This shall be elaborated in the section that follows.

\subsubsection{Various equations-of-state}

At finite cosmological constant, the Friedmann equations read \cite{0904.1595}
\begin{eqnarray}
\frac{\ddot{a}}{a} &=& \frac{2}{3\, \lambda -1} \left[\frac{\Lambda_{\omega}}{2} - \frac{4\, \pi\, G_N}{3} \left(\rho + 3 p\right)- \frac{K^2}{2\, \Lambda_{\omega}\, a^4}\right].
\end{eqnarray}
With continuity equation, this equation becomes
\begin{eqnarray}
\dot{a}^2 = \frac{\Lambda_{\omega}}{3\, \lambda-1}\, a^2 + \frac{16\, \pi\, G_N}{3(3\, \lambda-1)} \rho_0\, a^{-(1+3\omega)} - \frac{2 K}{(3\, \lambda -1)} + \dfrac{\kappa^2}{\Lambda_{\omega}(3 \lambda -1)}\, a^{-2}. \label{eq:bdHRGadd}
\end{eqnarray}
Due to various equations-of-state, this differential equation can be solved. 
\begin{itemize}
\item At $\omega=0$, Eq. (\ref{eq:bdHRGadd}) can be reduced to
\begin{eqnarray}
\dot{a}^2 = \frac{\Lambda_{\omega}}{3\, \lambda-1}\, a^2 + \frac{16\, \pi\, G_N}{3(3\, \lambda-1)} \rho_0\, a^{-1} - \frac{2 K}{(3\, \lambda -1)} + \dfrac{\kappa^2}{\Lambda_{\omega}(3 \lambda -1)}\, a^{-2}. \label{eq:dtldt1}
\end{eqnarray}
Then, the solution reads
\begin{eqnarray}
t &=& \frac{2}{3}\, D_1^{-2}\, \left(D_1\, a - 2 F_1\right)\, \left(D_1\, a + F_1\right)^{1/2},  \label{eq:bHLG1}
\end{eqnarray}
where $D_1=16\, \pi\, G_N\, \rho_0/[3(3\, \lambda-1)]$ and $F_1=K^2/[\Lambda_{\omega} (3\, \lambda-1)]$.

\item At $\omega=1/3$, we obtain
\begin{eqnarray}
\dot{a}^2 = \frac{\Lambda_{\omega}}{3\, \lambda-1}\, a^2 + \left(\frac{16\, \pi\, G_N}{3(3\, \lambda-1)} \rho_0 + \frac{K^2}{\Lambda_{\omega} (3\, \lambda-1)}\right)\, a^{-2} - \frac{2 K}{3 \lambda-1}, \label{eq:dtldt2}
\end{eqnarray}
which can be solved as follows.
\begin{eqnarray}
t =\frac{1}{2\, \sqrt{D_2}} \left[ \ln \left(a^2 + \frac{G_2}{2 D_2} + \sqrt{a^4 + \frac{F_2}{D_2}+\frac{G_2}{D_2}a^2} \right) \right], \label{eq:dtl2}
\end{eqnarray}
where $D_2=\Lambda_{\omega}/(3\, \lambda-1)$ and $F_2= 16\, \pi\, G_N\, \rho_0/[3(3\, \lambda-1)]  + K^2/[\Lambda_{\omega} (3\, \lambda-1)]$ an d $G_2 = - \frac{2K}{3 \lambda - 1}$. 

\item At $\omega=-1/3$, it results is
\begin{eqnarray}
\dot{a}^2 = \frac{\Lambda_{\omega}}{3\, \lambda-1}\, a^2  + \frac{\kappa^2}{\Lambda_{\omega} (3\, \lambda-1)}\, a^{-2} + \left[\frac{(16πG_N)}{3(3 \lambda-1)} \rho_\circ - \frac{2κ}{(3 \lambda -1)} \right], \label{eq:dtldt3}
\end{eqnarray}
with the solution
\begin{eqnarray}
t = \frac{1}{2\, \sqrt{D_2}} \left[ \ln \left(a^2 + \frac{G_2 + F_2 - F_3}{2D_2} + \sqrt{a^4 + \frac{F_3}{D_2} + \frac{G_2 + F_2 - F_3}{D_2} a^2} \right) \right], \label{eq:dtl3}
\end{eqnarray}
where $F_3=K^2/[\Lambda_{\omega} (3\, \lambda-1)]$. 

\item At $\omega=-1$, we get
\begin{eqnarray}
\dot{a}^2 = \left(\frac{\Lambda_{\omega}}{3\, \lambda-1} + \frac{16\, \pi\, G_N}{3(3\, \lambda-1)} \rho_0\right)\, a^2 + \frac{\kappa^2}{\Lambda_{\omega} (3\, \lambda-1)}\, a^{-2} - \frac{2K}{3 \lambda-1}, \label{eq:dtldt4}
\end{eqnarray}
with the solution
\begin{eqnarray}
t = \frac{1}{2\, \sqrt{D_2+F_2-F_3}} \left[\ln \left(a^2+\frac{G_2}{2(D_2+F_2-F_3)}+\sqrt{a^4+\frac{F_3}{D_2+F_2-F_3} + \frac{G_2}{D_2+F_2-F_3}a^2} \right) \right], \label{eq:dtl4}
\end{eqnarray} 

\item For the generalized Chaplygin gas \cite{chplyginREF}, it becomes
\begin{eqnarray}
\dot{a}^2 &=& \left( \frac{\Lambda_\omega}{3 \lambda-1} + \frac{16\, \pi\, G_N A^{\frac{1}{1+\alpha}}}{3(3\, \lambda -1)}\, a^2 +  \frac{K^2}{\Lambda_{\omega} (3\, \lambda-1)}\, a^{-2} - \frac{2K}{3 \lambda-1} \right), \label{eq:dtldt5} \\
t &=& \frac{1}{2\, \sqrt{D_2+D_5}}  \left[\ln \left(a^2+\frac{G_2}{2(D_2+D_5)}+\sqrt{a^4+\frac{F_3}{D_2+D_5}+\frac{G_2}{D_2+D_5}a^2} \right) \right], \label{eq:dtl5}
\end{eqnarray}
where  $D_5= 16\, \pi\, G_N\, A^{1/(1+\alpha)}/[3(3\, \lambda -1)]$.

In a matter-dominated universe, i.e. $p=0$ or $A=0$, then $\rho=B^{1/(1+\alpha)}\, a^{-3}$ and
\begin{eqnarray}
\dot{a}^2 = \frac{\Lambda_{\omega}}{3\, \lambda-1}\, a^2 + \frac{16\, \pi\, G_N B^{\frac{1}{1+ \alpha}}}{3(3\, \lambda-1)} a^{-1} + \frac{K^2}{\Lambda_{\omega} (3\, \lambda-1)}\, a^{-2} - \frac{2K}{3 \lambda-1}.
\end{eqnarray}
Similar to Eq. (\ref{eq:dtldt1}), $\Lambda_{\omega}$ is assumed to be negligibly small 
\begin{eqnarray}
t &=& \frac{1}{3}\, D_6^{-2}\, \left(2 D_6\, a - F_1\right)^{1/2}\, \left(D_6\, a + F_1\right),  \label{eq:bHLG1}
\end{eqnarray}
where $D_6=8\, \pi\, G_N\, B^{1/(1+\alpha)}/[3(3\, \lambda-1)]$.

\item For cosmological geometry filled with QCD matter, the EoS can be deduced from the recent lattice QCD simulations \cite{fodor2013}. As given in Fig. \ref{fig:peqcd}, $w$ ranges between $\sim 1/6$ and $\sim 1/4$.  For simplicity, the second term in Eq. (\ref{eq:bdHRGadd})  can be approximated,  
\begin{eqnarray}
\ddot{a} &\approx &  - \left[\frac{2\, \pi\, G_N}{(3\, \lambda-1)} \rho_0\, + \frac{K^2}{\Lambda_{\omega} (3\, \lambda-1)}\right]\, a^{-3} + \frac{\Lambda_{\omega}}{3\, \lambda-1}\, a. 
\end{eqnarray}
Similar to Eq. (\ref{eq:dtl2}), the solution reads 
\begin{eqnarray}
t &=&\frac{{\cal B}_7\, \log\left[2\left(\sqrt{D_2}\; {\cal B}_7 + D_2\, a^2\right)\right]}{2\, \sqrt{D_2}\; {\cal B}_7}, \label{eq:dtl2}
\end{eqnarray}
where ${\cal B}_7=\left[D_2\, a^4 - F_7\right]^{1/2}$ and $D_2=\Lambda_{\omega}/(3\, \lambda-1)$ and $F_7=2\, \pi\, G_N\,  \rho_0/(3\, \lambda-1) + K^2/[\Lambda_{\omega} (3\, \lambda-1)]$. 

\end{itemize}

\section{Results and Discussion}
\label{sec:disc}

Based on studying a $\lambda$-dependent version of the Friedmann equations, it was concluded that HLG becomes an attractive gravity theory if the generalized Wheeler-DeWitt metric ${\cal G}^{ijkl}=(g^{i\, k}\, g^{j\, l} + g^{i\, l}\, g^{j\, k})/2$ has an indefinite signature. But, it becomes a repulsive gravity theory when the metric gets a positive sign. These results are also found in our calculations. The various equations-of-state lead to different behavior of the scalar factor  as an example about the cosmological quantities.

In Fig. \ref{fig:1}, the cosmic time ($t$) is given in dependence on the scale factor ($a$). In these calculations all parameters are kept fixed, $k=G=c=\rho_0=1$ and $\mu=0.5$, $\lambda=0.9$ and $\Lambda_{\omega}=0.1$. Correspondingly, neither $t$ nor $a$ is given in physical units. We compare between  HLG with non-detailed and detailed balance conditions and GR gravity for various equations-of-state. It is worthwhile to notice that FLRW which is characterized by GR gravity is non-singular for four types of equations of state; $\omega=1/3$ (a), $\omega=0$ (b), $\omega=-1/3$ (c) and $\omega=-1$ (d). 

The singularity in the FLRW cosmology, which can be characterized by HLG with non-detailed and detailed balance conditions, varies due to the equations of state. Such conclusions are only valid for the given parameters. Furthermore, the dependence of $t$ on $a$ apparently varies with the equation of state, as well. Again, such conclusions are explicitly valid for the given parameters. In other words, absence or presence of singular solutions and even the dependence of $t$ on $a$ strongly depend on the equation of state and the gravity itself. The non-singular solutions is obvious, as $\Lambda_{\omega}$ is finite.

The top-left panel shows the results at $\omega=1/3$ (a), i.e. radiation-dominated  (perfect fluid approximation) FLRW universe. The HLG with non-detailed balance conditions, Eq. (\ref{eq:ndwp13}) leads to a very rapid increase in $a$, almost a parabolic shape. At a certain value of $t$, the scale factor became $t$-independent, i.e. $a$ diverges. The results from HLG with detailed balance conditions, Eq. (\ref{eq:dtl2}), and GR, Eq. (\ref{eq:w13std}), are also compared with.  

The bottom-left panel shows the results at $\omega=0$ (b), i.e. matter-dominated  (dust approximation) FLRW universe. Again, the HLG with non-detailed,  Eq. (\ref{eq:add0a}), and detailed balance conditions, Eq. (\ref{eq:bHLG1}), lead to a rapid increase in $a$; an almost a linear dependence. Also here, the GR gravity, Eq. (\ref{eq:EGw1}), has a non-singular solution. The bottom-right panel presents the results at $\omega=-1$ (d), i.e. vacuum or dark energy or finite cosmological constant. We find that the validity of  HLG with non-detailed balance conditions is very limited. Also, HLG with detailed balance conditions shows a very rapid change in $a$ against $t$. The GR gravity results in an inflationary phase followed by a slower expansion.

Despite the non-singular solutions, we notice that the expansion depicted in the right-hand panel is faster than the one illustrated in the left-hand panel. Again, the universe seems to start up its evolution from a very large scale factor. 

Figure \ref{fig:2} presents the results from Chaplygin (a) and QCD (b) equations of states. The results of the Chaplygin gas strongly depends on the parameter $A$. The HLG with  detailed balance conditions results in an almost linear dependence of $a$ and $t$. The non-singular solutions appear in both HLG approaches and equations of state. We conclude that the difference between the Chaplygin and QCD results is due to non-baryonic and baryonic equations-of-state, respectively.

While the present article was under review, we have completely conducted and published a research paper in which the proposed solutions are confronted to recent PLANCK and BICEPII observations \cite{Tawfik:2016dv}. With single scalar field potentials describing power-law and minimal-supersymmetrically extended inflation, we have derived possible modifications in the Friedmann equations. For various EoS, the dependences of the tensorial and spectral density fluctuations (and their ratio) on the inflation field are characterized. The tensorial-to-spectral density fluctuations are calculated with varying spectral index. It was found that, they decrease when moving from HLG with non-detailed balance conditions, to Friedmann gravity, to HLG without the projectibility conditions, and to  HLG with detailed balance conditions. Such a pattern remains valid for various EoS and different inflation potential models. The calculations fit well with the recent PLANCK observations \cite{planck2015a,planck2015b}.

Even, when studying the consequences of the quantum fluctuations on our understanding of Landau-Raychaudhuri equations, we found that cosmic EoS play an essential role \cite{TD-INJP1}.

\section{Conclusions}
\label{sec:conc}

A theory for quantum gravity with an anisotropic scaling in UV based on a scalar field theory was proposed by Horava, who was inspired by Lifshitz theory for the quantum critical phenomena in condensed matter physics. Horava benefited from the success of QFT in describing all forces (except gravity) and the standard model for the particle physics. We study the impacts of various equations-of-state on FLRW cosmology based on Horava-Lifshitz gravity. We also implement the so-called  HLG with detailed balance conditions, which was thought as a technical trick to reduce couplings in the theory. From the possible connection between detailed balance conditions and the entropic origin of the gravity, the constructed potential can be used instead of adopting an effective field theory. Both are compared with the GR gravity.

In this work, we have studied the dependence of scale factor ($a$) and cosmic time ($t$) in the context of Horava-Lifshitz gravity in early universe by using different equations-of-state. We have compared the results with the GR gravity.  Remarkable differences from what is predicted by GR are found. Also, a noticeable dependence on the equations of state is observed in singular and non-singular Big Bang. These observations are precautionarily corresponding to the fixed parameters $k=G=c=\rho_0=1$ and $\mu=0.5$, $\lambda=0.9$ and $\Lambda_{\omega}=0.1$. The results presented in present paper are strongly depending on these parameters, for instance, the non-singular solutions can be to a large extend understood due to finite cosmological constant, $\Lambda_{\omega}$.

The present work presents a systematic analysis for the equations of state characterizing the cosmic geometry and proposes a link between the conventional gravity by GR and its Horava-Lifshitz counterparts. The latter are power-counting renormalizable quantum theories which assure causal dynamical triangulations, renormalization group approaches based on asymptotic safety and symmetries of GR. The Horava-Lifshitz gravity introduces a new set of symmetries imposing invariance under foliation-preserving diffeomorphisms. Various types of equations of state are implemented. 

In future works, we intend to extend this study to cover various topics, such as comparison with other theories such as the gravity's rainbow \cite{ReFf1,ReFf2,ReFf3}, and determining the possible impacts of the generalized uncertainty principle \cite{RefGUP1,RefGUP2,RefGUP3}. Also, we plan to study third quantization of this theory \cite{RefThird1,RefThird2}, such as in a simple cosmological model \cite{RefThirdModel}.

We conclude that HLG can explain various epochs in the early universe.  Furthermore, HLG might be able to reproduce the entire cosmic history with and without singular Big Bang. It intends to study the stability of the given FLRW solutions and then analyse the universe evolution scenarios, especially that HLG is conjectured to be a candidate for a quantum field theory of gravity, despite its yet-unresolved problems.

\appendix

\section{Appendix A: Einstein Gravity}

For EoS $p=\omega\, \rho$ and Friedmann equations with finite cosmological constant, the continuity equation reads
\begin{eqnarray}
\dot{a}^2 =  \frac{8\, \pi\, G}{3} \rho_0\, a^{-(1+3\, \omega)} + \frac{1}{3}\, \Lambda_{\omega}\, a^2. \label{eq:egDE2}
\end{eqnarray} 
Accordingly, we get following solutions
\begin{eqnarray}
\mathrm{at}\; \omega=0,\;\;\; t &=& \frac{2 {\cal C}_1 \log\left[2\left(\sqrt{B}{\cal C}_1 + B a^{3/2}\right)\right]}{3\, \sqrt{B}\, {\cal C}_1} ,\label{eq:EGw1} \\
\mathrm{at}\; \omega=\frac{1}{3},\;\;\; t &=& \frac{1}{\sqrt{B}} \ln \left[a^2 - \dfrac{2A}{B} \right], \label{eq:w13std} \\
\mathrm{at}\; \omega=-\frac{1}{3},\;\;\; a(t) &=& c_1 \exp\left(\sqrt{B}\, t\right) + c_2 \exp\left(-\sqrt{B}\, t\right), \\
\mathrm{at}\; \omega=-1:\;\;\; a(t) &=& c_3 \exp\left(\sqrt{C}\, t\right) + c_4 \exp\left(-\sqrt{C}\, t\right), \\
\mathrm{and}\, \mathrm{for}\, \mathrm{generalized}\, \mathrm{Chaplugin}\, \mathrm{gas},\; a(t) &=& c_5 \exp\left(\sqrt{D}\, t\right) + c_6 \exp\left(-\sqrt{D}\, t\right),
\end{eqnarray}
where ${\cal C}_1=\sqrt{B\, a^3 - 2 A}$, $A=-4\, \pi\, G\, \rho_0/3$, $B=\Lambda_{\omega}/3$, and $C=-2A+B$, $D=8\, \pi\, G_N\, A^{\frac{1}{1+\alpha}}/3$. $c_1 \cdots c_6$ are  integration {\it arbitrary} constants.

\section{Appendix B: QCD Equation of State}

From the recent lattice QCD simulations \cite{fodor2013}, we illustrate in Fig. \ref{fig:peqcd}, the pressure in dependence on the energy density. As discussed in Ref. \cite{Tawfik:2015kwa}, the ultimate goal of the high-energy experiments is first-principle determination of the underlying dynamics in the strongly interacting matter. Determining the thermodynamical quantities, which are well-known tools to describe nature, degrees of freedom, decomposition, size and even the overall dynamics controlling evolution of the medium from which they are originating, is therefore very essential. For the thermodynamical pressure, an approximate attempt utilizing the higher-order moments of the particle multiplicity seems to be promising \cite{Tawfik:2015kwa}.  On the other hand, determining the thermodynamical energy density and even relating the Bjorken energy density to the lattice energy density depends on lattice QCD at finite baryon chemical potential and first-principle estimation of the formation time of the quark-gluon plasma (QGP). The energy density can be deduced from the derivative of the free energy with respect to inverse temperature. This would explain the need to implement another variable different than the one responsible for the derivatives of the high-order moments. 

In Fig. \ref{fig:peqcd}, the lattice QCD results on pressure vs. energy density are fitted as follows. 
\begin{itemize}
\item Hadron: 
\begin{eqnarray}
p &=& (0.157\pm 0.0007)\, \rho^{1.008\pm 0.07} \approx \frac{1}{6}\, \rho.
\end{eqnarray}
\item Parton: 
\begin{eqnarray}
p &=& (0.222\pm 0.002)\, \rho^{1.069\pm 0.0024} \approx \frac{1}{5}\, \rho.
\end{eqnarray}
\item Hadron-parton: 
\begin{eqnarray}
p &=& -0.01757 + (0.2413\pm 0.0007)\, \rho^{1.053\pm 0.008} \approx \frac{1}{4}\, \rho.
\end{eqnarray}
\end{itemize}



\begin{figure}[hbt]
\centering{
\includegraphics[width=5.cm,angle=-90]{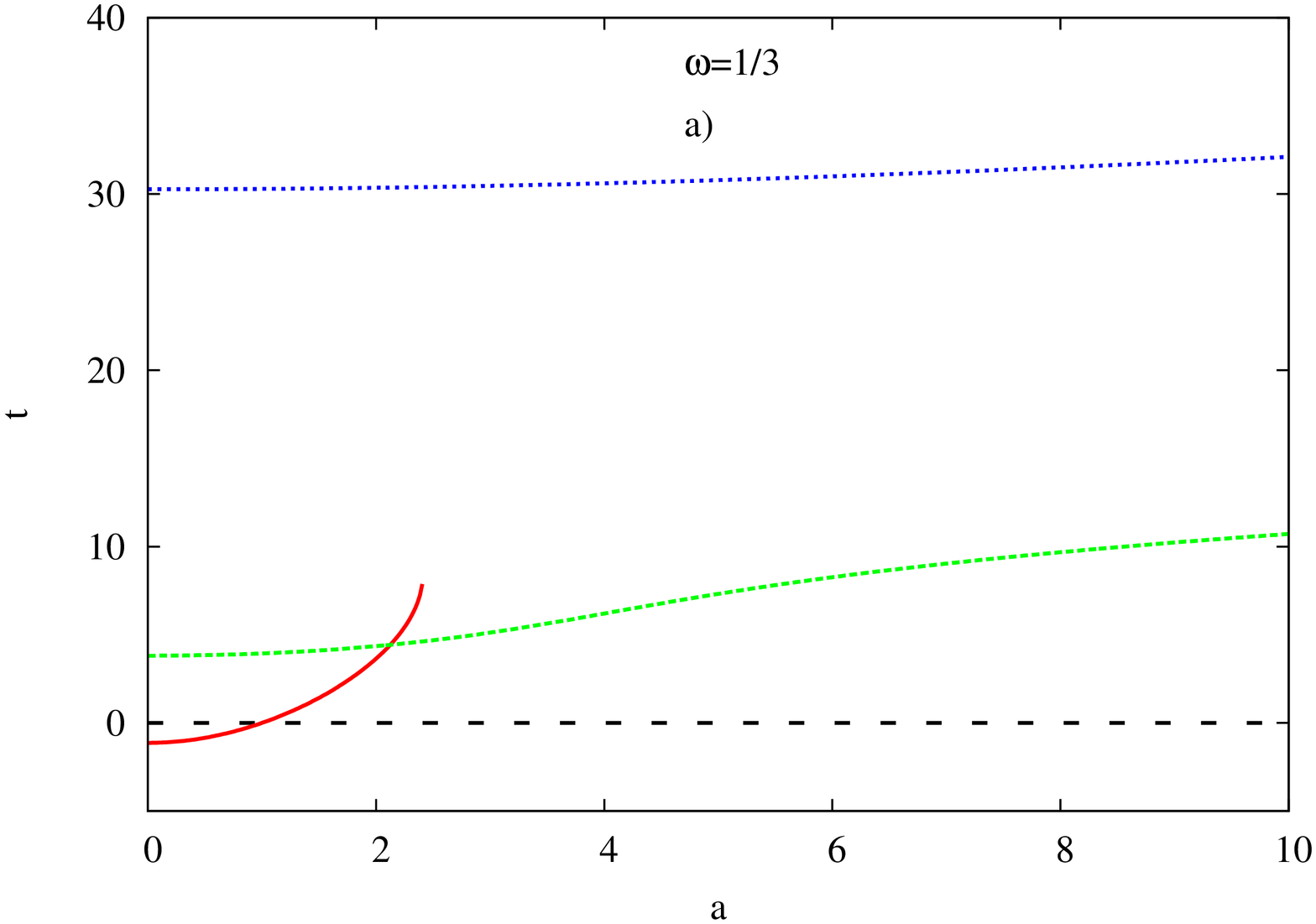}
\includegraphics[width=5.cm,angle=-90]{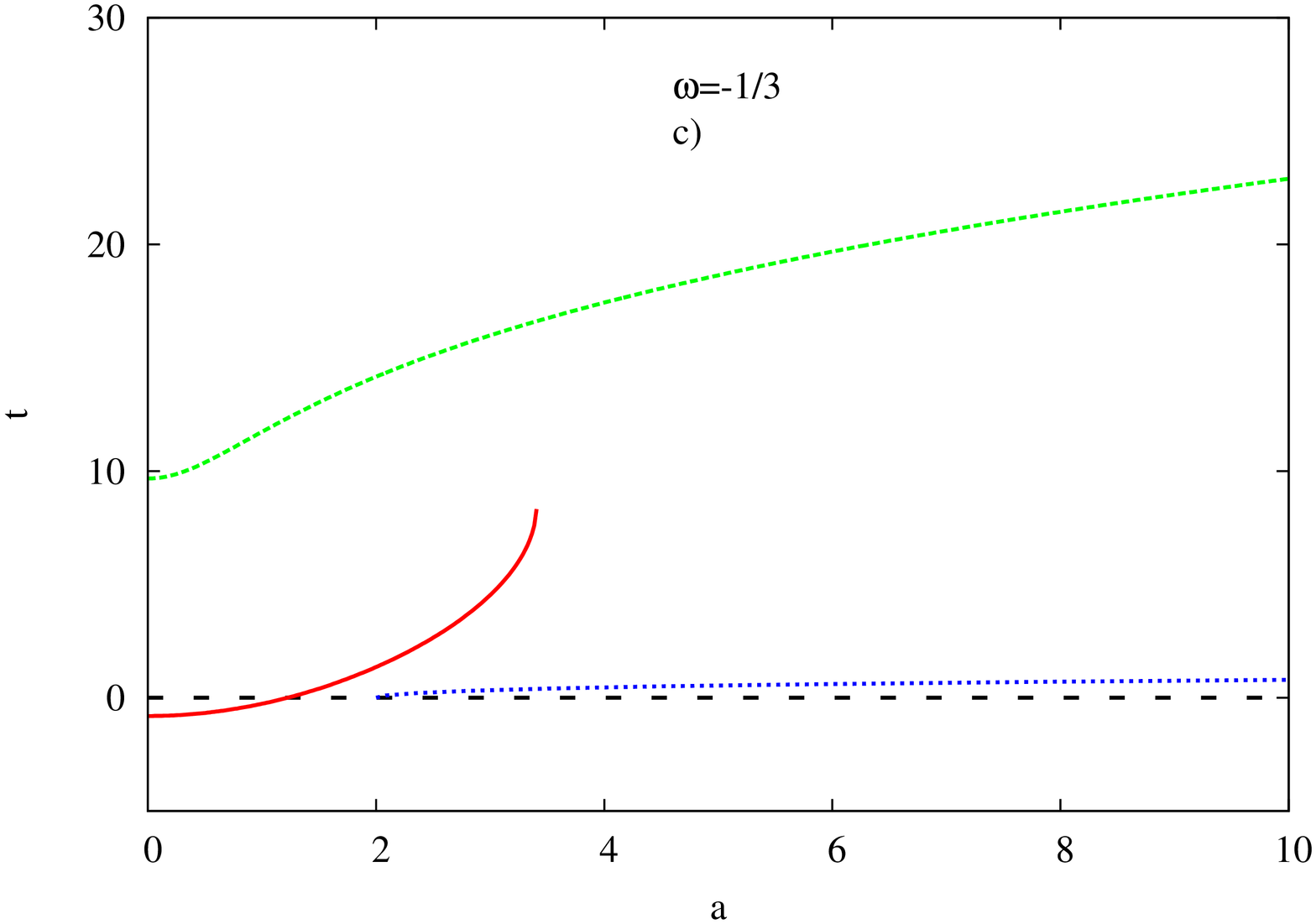} \\
\includegraphics[width=5.cm,angle=-90]{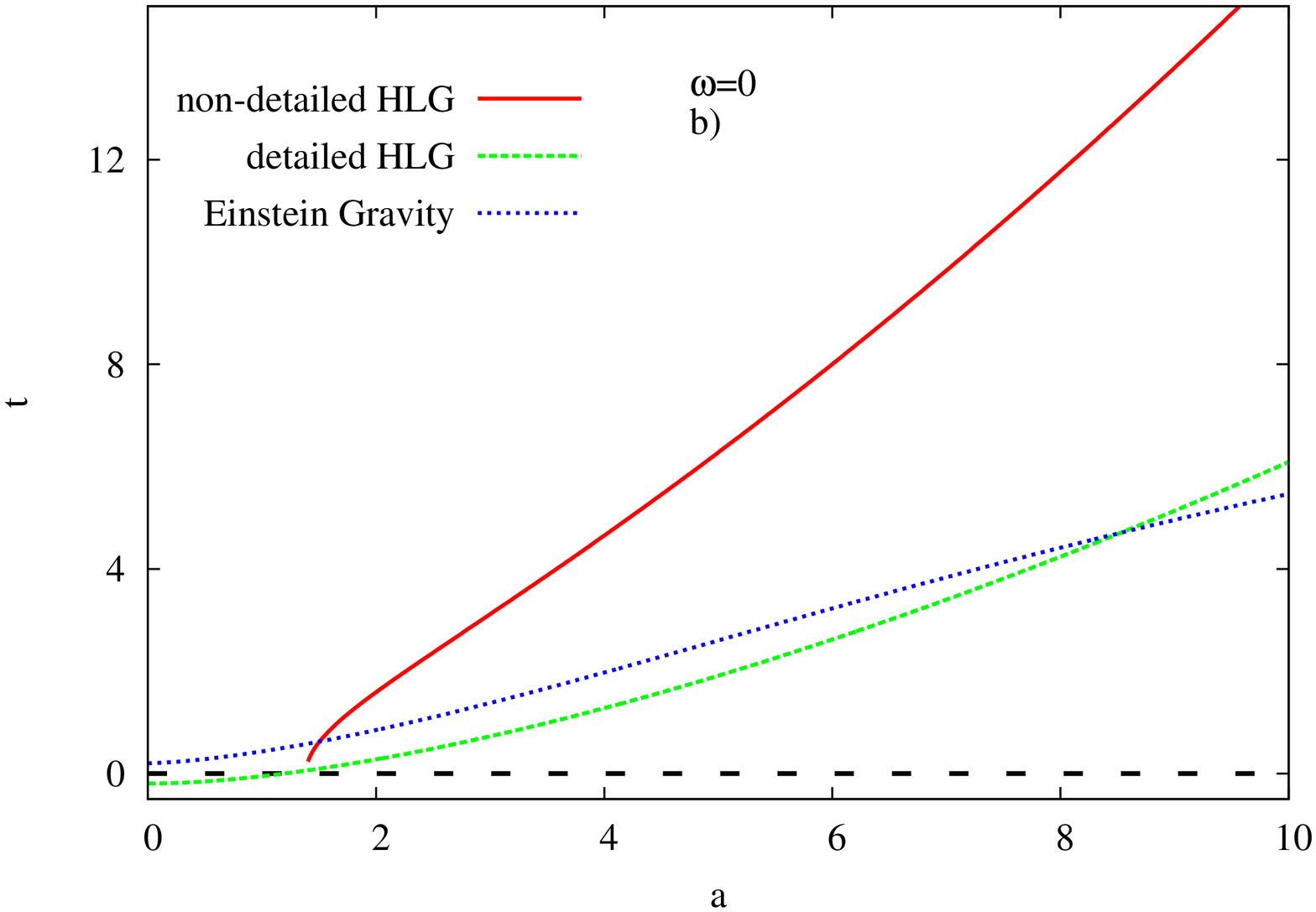}
\includegraphics[width=5.cm,angle=-90]{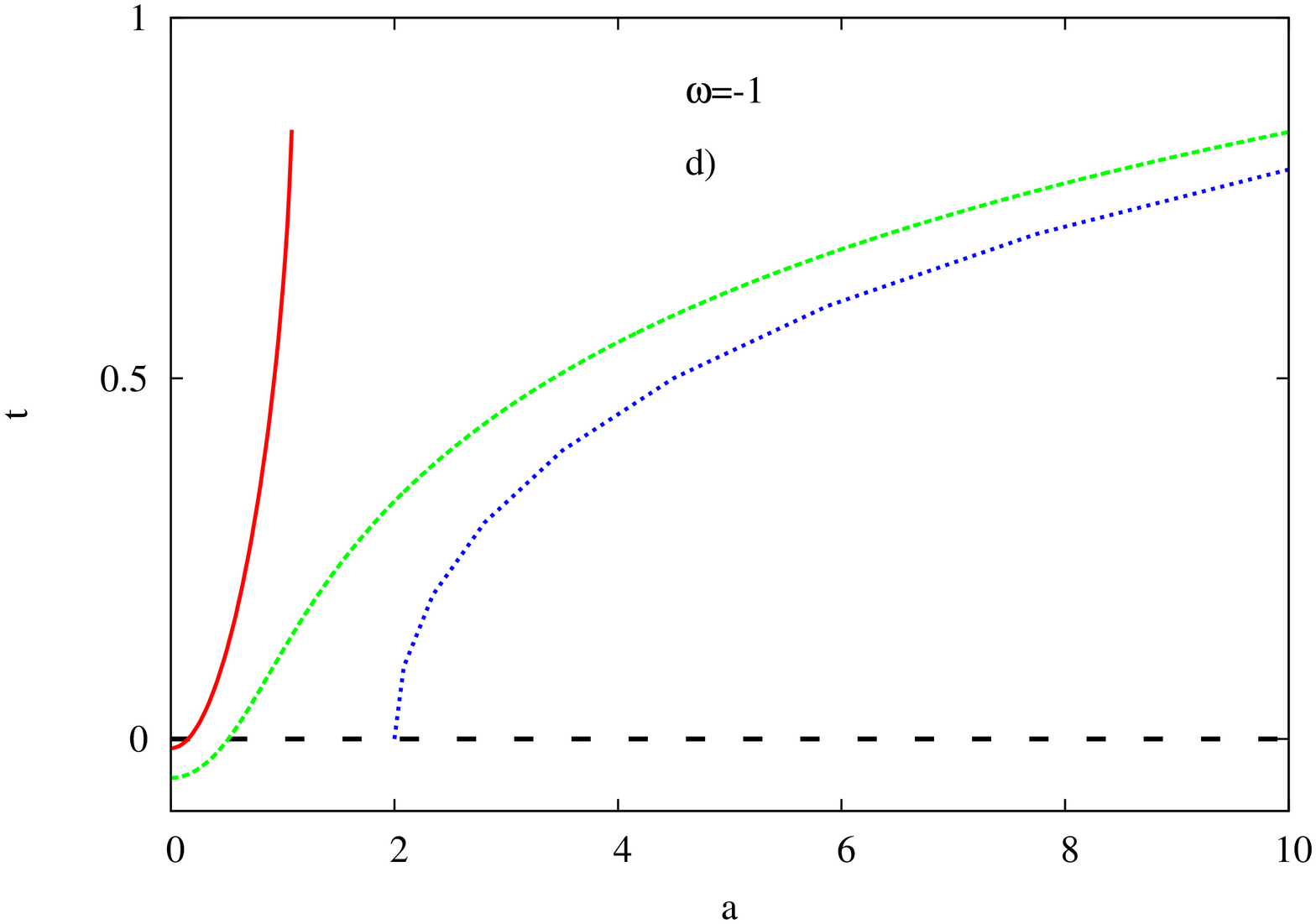} 
\caption{The cosmic time is given in dependence on the scale factor. We compare between non-detailed and detailed balance HLG and GR (Einstein) gravity using various equations of state, $\omega=1/3$ (a), $\omega=0$ (b), $\omega=-1/3$ (c) and $\omega=-1$ (d). All parameters are kept fixed, $k=G=c=\rho_0=1$ and $\mu=0.5$, $\lambda=0.9$ and $\Lambda_{\omega}=0.1$. Correspondingly, neither $t$ nor $a$ has physical units. \label{fig:1} }
}
\end{figure}

\begin{figure}[hbt]
\centering{
\includegraphics[width=5.cm,angle=-90]{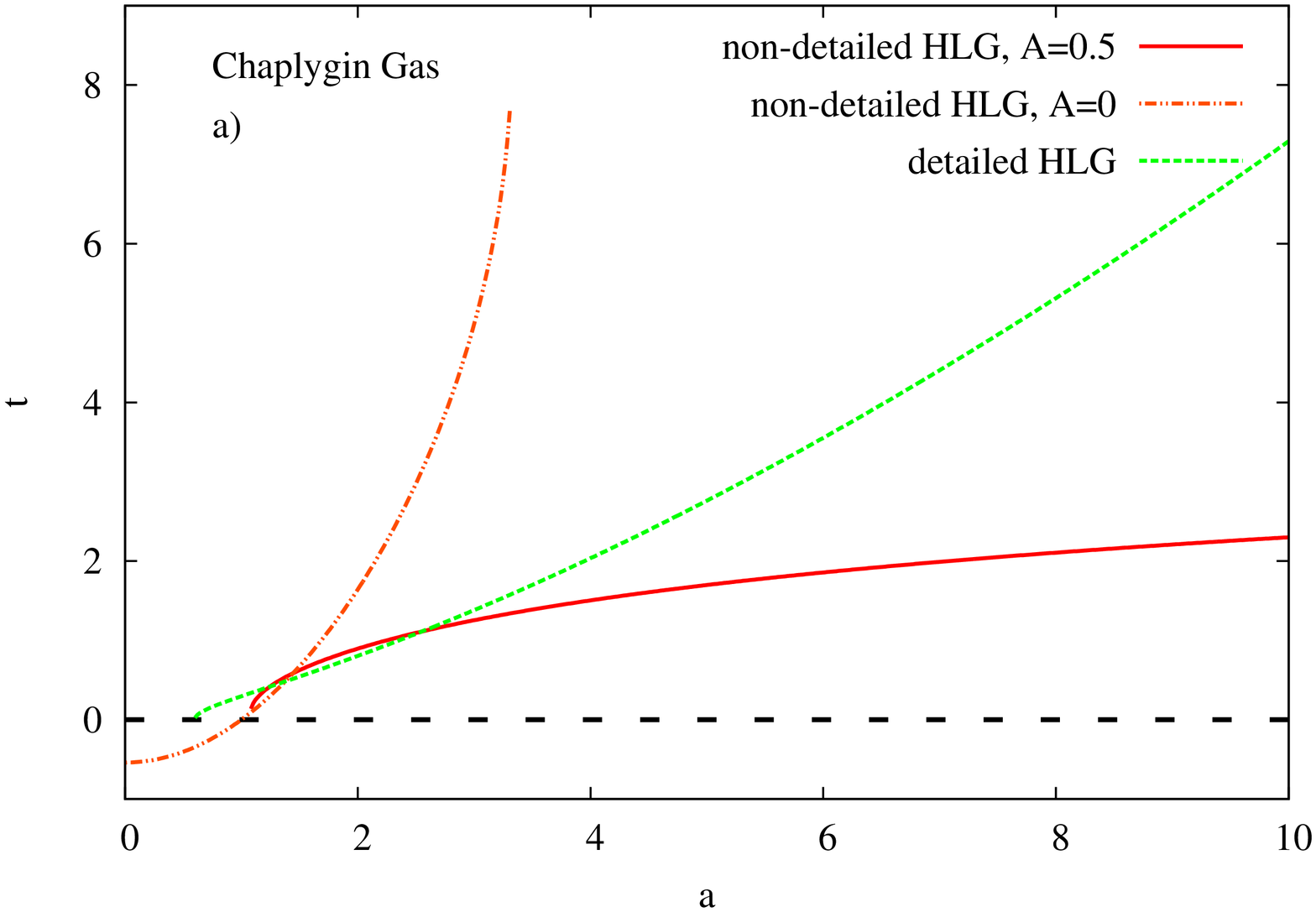}
\includegraphics[width=5.cm,angle=-90]{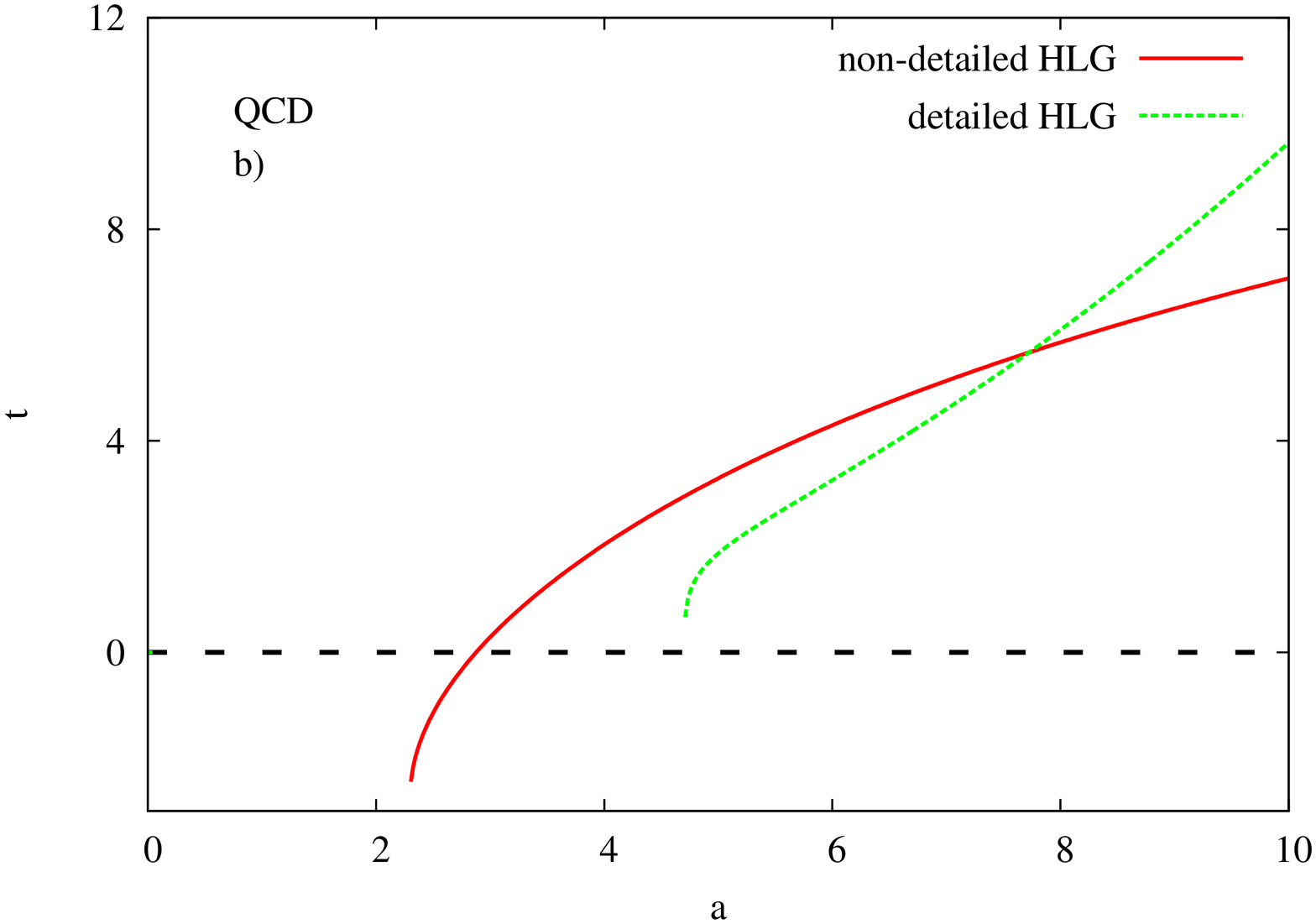}
\caption{As in Fig. \ref{fig:1} but for Chaplygin gas (a) and QCD (b) equations of state. \label{fig:2} }
}
\end{figure}

\begin{figure}[hbt]
\centering{
\includegraphics[width=7.cm,angle=-90]{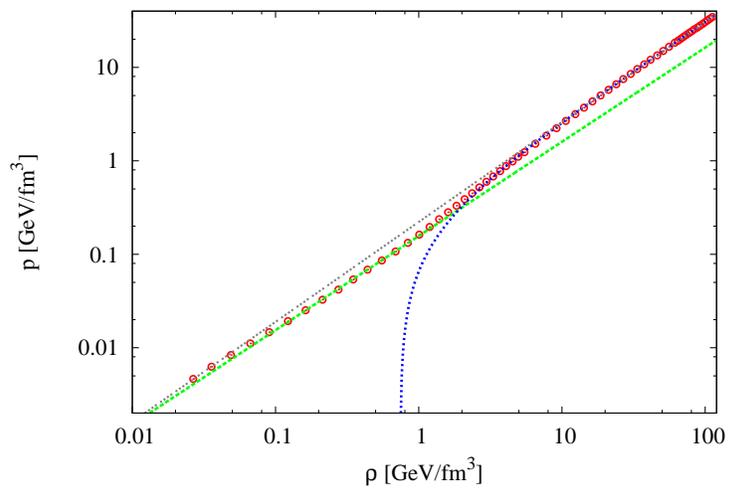}
\caption{In units of GeV/fm$^3$, the pressure is given as function of the energy density (symbols). The curves represent fits for hadron (long-dashed) and parton phase (dashed curves), separately. Both phases are fitted by the dotted curve. \label{fig:peqcd} }
}
\end{figure}

\end{document}